\documentclass[sigconf, nonacm]{acmart}
\usepackage[utf8]{inputenc}

\hypersetup{
    colorlinks,
    linkcolor={red!50!black},
    citecolor={blue!50!black},
    urlcolor={blue!80!black}
}

\newcommand\blfootnote[1]{%
  \begingroup
  \renewcommand\thefootnote{}\footnote{#1}%
  \addtocounter{footnote}{-1}%
  \endgroup
}

\usepackage{graphicx}

\newcommand{\Cref}[1]{Section~\ref{#1}}

\begin{document}

\title[Rescuing Counterspeech]{Rescuing Counterspeech: A Bridging-Based Approach\\to Combating Misinformation}
\author{Kenny Peng}\affiliation{\institution{Cornell University}\department{Cornell Tech; Department of Computer Science}\city{New York City}\country{United States}}
\author{James Grimmelmann}\affiliation{\institution{Cornell University}\department{Cornell Tech; Law School}\city{New York City}\country{United States}}

\date{\today}

\begin{abstract}
    Social media has a misinformation problem, and counterspeech---fighting bad speech with more speech---has been an ineffective solution. Here, we argue that \textit{bridging}-based ranking---an algorithmic approach to promoting content favored by users of diverse viewpoints---is a promising approach to helping counterspeech combat misinformation. By identifying counterspeech that is favored both by users who are inclined to agree and by users who are inclined to disagree with a piece of misinformation, bridging promotes counterspeech that persuades the users most likely to believe the misinformation. Furthermore, this algorithmic approach leverages crowd-sourced votes, shifting discretion from platforms back to users and enabling counterspeech at the speed and scale required to combat misinformation online. Bridging is respectful of users' autonomy and encourages broad participation in healthy exchanges; it offers a way for the free speech tradition to persist in modern speech environments.
    \blfootnote{We thank members of the AI, Policy, and Practice (AIPP) working group and the Digital Life Initiative (DLI) for thoughtful discussion and feedback.}
\end{abstract}

\maketitle

\section{Introduction}

Misinformation on social media threatens public health, people's reputation and safety, and healthy political discourse. Meanwhile, the long-preferred solution of combating false speech with \textit{counterspeech} appears ineffective. Misinformation is published with unprecedented speed and scale; filter bubbles prevent readers of misinformation from seeing corrections; fact checks often fail to change minds. So despite objections to speech regulation, the scale is tipping in this direction. Platforms have taken an increasingly proactive approach to content moderation in the past two decades, removing posts and blocking content as a response to harmful speech \cite{goldman2021content}.

Deciding what speech ought and ought not to be restricted is a difficult question---one we do not intend to address here. Even if there is disagreement on the details, there is a broad consensus that some material will always be over the line and that platforms ought to take it down: child sexual abuse material (CSAM), wholesale copyright infringement, cryptocurrency scams, credible individually targeted threats of violence, and so on. Content removals are not going away entirely---but perhaps they do not need to be used as widely as they currently are.

We are interested in the possibility of ``rescuing'' counterspeech for the purpose of combating false speech. Even as counterspeech appears particularly ineffective on social media, we suggest that many of the challenges that prevent counterspeech from achieving its purpose can be remedied online.  

This is not a new ambition. Since the rise of the Internet, scholars and activists have celebrated its ability to help protesters and other counter-speakers get their messages out \cite{kreimer2001technologies}. Counterspeech online benefits from the same reduced barriers to access as the speech it responds to, and social media in particular has made publishing counterspeech easier than ever. Whereas before, speech could only be widely disseminated by those with access to the right channels (newspapers, radio, television), now, anyone can reply on social media to even the most powerful political actors. 

But even if counterspeech is more accessible today, it is not effective. Counterspeech on the Internet faces  two related challenges. First, it has to be \emph{seen}: in an environment of extreme informational abundance, there is no guarantee that the people who were exposed to speech will see the reply. Second, it has to be \emph{believed}: in an environment of social suspicion and partisan polarization, even people who see counterspeech may disregard it because it contradicts their existing beliefs or because it comes from a source they are already inclined to distrust. The widely used phrases ``filter bubble'' and ``echo chamber'' capture how these two dynamics interrelate: people mostly \emph{see} messages coming from groups they already agree with, and mostly \emph{believe} messages coming from within those groups. The fact that the counterspeech exists online does little to change anyone's mind.

Here, we argue for a ``bridging-based'' approach \citep{ovadya2023bridging} to promoting counterspeech. Under \textit{bridging}, counterspeech is promoted on the basis of approval from users of diverse viewpoints. While this is not the most obviously direct way to correct misinformation, it is desirable by many measures. Chief among them is effectiveness. In particular, we argue that in environments of ideological polarization, counterspeech can effectively correct misinformation \textit{if and only if} it is bridging. Consider a piece of misinformation that group $A$ is inclined to believe and group $B$ is inclined to disbelieve. Counterspeech that wins group $A$'s approval but not group $B$'s is unlikely to be corrective in its \emph{content}; it will reinforce group $A$'s belief in the original false or misleading claim rather than challenging it. Counterspeech that wins group $B$'s approval but not group $A$'s is unlikely to be corrective in its \emph{effect}; the information in it will not change the minds of members of group $A$. Counterspeech that is approved by both groups is both likely to correct the original claim (because group $B$ approves it) and to present information that believers of the misinformation find compelling (because group $A$ approves it).

\Cref{sec:background} provides the relevant background in how free speech theories inform evaluation of counterspeech, and what counterspeech must do to succeed in the modern speech environment. While the efficacy of counterspeech in combating misinformation is a primary concern, it is not the only desiderata by which to evaluate a ``counterspeech mechanism'' (a way to promote counterspeech). Others include respect for individuals' autonomy and distrust of concentrated power. Bridging-based counterspeech is an appealing approach under many theories of free speech; it inherits many of the familiar desirable properties of counterspeech, while adding a few new ones. For example, a primary concern in the search-for-truth justification of free speech is that regulation is likely to bend to private or government interests; bridging relieves this concern by placing discretion collectively with the users of a platform. Bridging-based counterspeech is also practically appealing. For example, any effective online counterspeech mechanism must operate at the speed and scale of social media. Bridging succeeds here because it is implemented by crowd-sourcing. 

Attempts to promote counterspeech are not new. Fact-checking organizations' raison d'\^{e}tre is to produce counterspeech to combat misinformation. Less directly, social media algorithms decide which replies or comments to display, typically to maximize metrics like engagement or relevance. Further afield, the Fairness Doctrine and right-of-reply laws promoted counterspeech by requiring that diverse viewpoints be expressed on broadcast television and radio. But these precedents---discussed in \Cref{sec:precedents}---faced severe theoretical, political, and practical challenges. For example, while the fairness doctrine prescribes a relatively simple objective (``show both sides''), adjudication of the rule in practice was nearly impossible, due to significant subjectivity. Bridging can help overcome this challenge since it is implemented by a concrete algorithm, established and calibrated without reference to any particular speech dispute.

At a high level, bridging is a promising approach because of two features: First, the bridging algorithm \emph{identifies} speech that effectively combats misinformation. Other recommendation algorithms can identify content that an individual user is most likely to find appealing, but the bridging approach surfaces consensus content that is especially likely to respond to mistaken user beliefs. Second, the \textit{implementation} of the bridging objective facilitates the large-scale promotion of counterspeech in a way that is in alignment with theoretical roles of counterspeech and practical challenges that have rendered other counterspeech mechanisms ineffective. 

\Cref{sec:bridging} makes both of these points. In this section, we describe the algorithmic implementation of bridging---which adapts collaborative filtering techniques used in recommender systems to identify bridging counterspeech. In particular, we describe the algorithm used by X's Community Notes program, a real-world implementation of the approach we describe here \citep{wojcik2022birdwatch}. We then analyze the efficacy of the Community Notes program thus far, presenting nascent but accumulating evidence suggesting its promise---but also weaknesses. Early evidence suggests that Community Notes succeeds at identifying counterspeech that users find persuasive, reduces how much misinformation is shared, and produces corrections that are accurate. However, there is also evidence that the current implementation is less effective at curtailing the early spread of misinformation, suggesting that the algorithm may require too many votes to reach consensus. There are also concerns about that the algorithm may be gameable in ways that could limit its efficacy. These considerations present important directions for future work in algorithm design.

Before we begin, a note on terminology. We use ``misinformation'' to refer to \emph{false or misleading} speech: i.e., speech that makes objectively untrue claims about factual matters, or speech that is literally true but causes readers to form false beliefs unless it is accompanied by additional context.  The term ``misinformation'' is controversial, but most of those controversies concern how much misinformation actually takes place, and how much of a problem it is. Those who object to the term most commonly object to proposed interventions---such as more aggressive content moderation---because they worry that these interventions are actually designed to promote one side of a controversial issue over the other. One of the virtues of a bridging-based approach is that it should be more appealing to people who do not believe that misinformation is a significant problem in the first place. It adds context to speech rather than removing it entirely, it gives platforms less discretion to pick sides in a dispute, and it promotes only those responses that genuinely add useful context, \emph{as agreed upon by members of the groups that see it}.

\section{Background}\label{sec:background}

\subsection{Theories of Speech and Counterspeech}

There are numerous theories of free speech. In this section, we canvas three families of theories: the search for truth, individual autonomy, and self-government. As we will see, the search for truth has by far the most to say about counterspeech, and it makes the most sense to evaluate counterspeech proposals from a search-for-truth perspective. Our aim in this section will be to show that effective counterspeech affirmatively achieves the goals of search-for-truth theories, and that other theories are generally in favor of it.

While they agree on core cases, each theory of free speech has its own rationale and its own explanation of how far that rationale extends. In general, however, these theories all approve of counterspeech. First, counterspeech is itself speech, so they approve of additional people exercising their rights to free speech. Second, counterspeech is often presented as an \emph{alternative} to more restrictive measures, as in the adage (paraphrasing Justice Brandeis) that ``the best remedy for bad speech is more speech.''  The idea here is that these theories all regard government censorship of speech, even ``bad'' speech, as something to be avoided if at all possible. To the extent that ``more'' speech (i.e., counterspeech) can avert the harms from ``bad'' speech, it makes these other restrictions unnecessary. Counterspeech is therefore a preferred substitute for other restrictive measures.

These critiques of government censorship can also extend to content moderation by platforms \citep{land2019}. Here, however, there is a deep ambiguity. On the one hand, when a platform chooses what content to block, this can look like censorship, which these theories disapprove of. On the other hand, when the platform chooses what content to carry, this can look like the platform's own exercise of speech, which these theories approve of. The former view suggests that government might need to regulate platforms to make them stop censoring speech they dislike; the latter view suggests that such regulations themselves are the censorship. The contrast between these views is on display in the dueling majority and dissenting opinions in the recent United States Supreme Court case of \textit{Moody v. NetChoice} \citep{moody2024}. To the extent that voluntarily applied bridging-based counterspeech can mitigate what platforms see as the harms of low-quality content, it provides a bridge between these two views of platform power.

In addition, note that each theory can also be viewed through a speaker-oriented lens---in which it asks how speech benefits speakers---or through a listener-oriented lens---in which it asks how speech benefits listeners \citep{grimmelmann2019listeners}. The listener-oriented lens is particularly useful for thinking about counterspeech, because it is speech and counterspeech's dueling effects on listeners that makes counterspeech of interest at all.

Finally, note that each theory comes with nuanced variations. We focus here only on broad families of free-speech theories to give a sense of the main arguments.

\paragraph{The Search for Truth.}

Counterspeech, especially as a way to mitigate misinformation, is most commonly situated in a marketplace-of-ideas theory of free speech, under which free speech protections are justified by the ``search for truth.'' This theory has played a central role in First Amendment jurisprudence, going back to Justice Holmes' famous dissent in \textit{Abrams v. United States}, in which he argued, ``the best test of truth is the power of the thought to get itself accepted in the competition of the market, and that truth is the only ground upon which their wishes safely can be carried out'' \citep{abrams}. A decade later, in \textit{Whitney v. California}, Justice Brandeis elaborated, ``If there be time to expose through discussion the falsehood and fallacies, to avert the evil by the processes of education, the remedy to be applied is more speech, not enforced silence'' \citep{whitney}. 

The search-for-truth theory is fundamentally empirical. It argues that the goal of free speech and free-speech law is to help society converge on true beliefs rather than false, and to exchange useful information rather than harmful. The preference for counterspeech therefore rests on two basic premises: (1) that censorship is bad for the search for truth, and (2) that counterspeech can effectively combat bad (e.g., false) speech in lieu of censorship.

On the first point, Eugene Volokh offers three standard reasons for why government censorship of speech harms the search for truth \citep{volokh2013}. First, government regulation of speech is more likely to be dictated by private or political interests, rather than a genuine interest in promoting truth. Allowing censorship to advance search-for-truth objectives opens the use of censorship towards other objectives. Second, the censorship of speech prevents ideas and facts from being re-evaluated and revised over time. What is considered true today may not be years from now, when new evidence emerges. History is ripe with examples where the conventional wisdom was later shown wrong. Third, censoring challenges to a truth can prevent acceptance of that truth, since skeptics will remain.  These reasons against censorship support a laissez-faire ``market'' approach, where the search for truth is best supported by the free, uncensored exchange of ideas, under which it can be ``accepted in the competition of the market.'' 

The second assumption of the marketplace theory is that counterspeech in fact succeeds in enabling the search for truth. In \textit{United States v. Alvarez}, Justice Kennedy wrote, ``The Government has not shown, and cannot show, why counterspeech would not suffice to achieve its interest. The facts of this case indicate that the dynamics of free speech, of counterspeech, of refutation, can overcome the lie'' \citep{alvarez2012}.
Many have noted that the marketplace theory is flawed in making this assumption---and that the modern speech environment especially does not operate at all like this ideal. Zeynep Tufekci writes that the marketplace theory is ``flatly belied by the virality of fake news'' \citep{tufekci2018}. Evelyn Douek observes that this change in calculus---where counterspeech is recognized as fallible---has led to ``a turn from First Amendment absolutism to the need to balance interests proportionally'' \cite{douek2021governing}. In the next section, we will more carefully lay out some reasons for why counterspeech fails in practice.

Evaluating whether counterspeech is effective in promoting truth requires a deeply listener-oriented analysis, because the question of whether false speech or true counterspeech is more persuasive is fundamentally a question of listener sociology (does the counterspeech \emph{reach} listeners?) and psychology (do they \emph{believe} it?). Indeed, counterspeech that falls in a forest with no one around to hear it is pointless in a search-for-truth theory. Counterspeech interventions are valuable to the extent that they are effective, which means to the extent that they persuade listeners exposed to the original harmful speech.

To summarize, under the marketplace theory, the efficacy of counterspeech is measured by its ability to facilitate the search of truth---in alignment with the goal of combating misinformation. The validity of the theory---in its prescriptive capacity---turns on counterspeech's efficacy.

\paragraph{Individual Autonomy.}

Whereas the search-for-truth theory is instrumentalist---speech is a means to the end of understanding the world---individual autonomy theories regard self-expression as a worthy end in itself. Every person has intrinsic dignity as a rational, creative, and social being, and to prohibit a person from expressing their ideas is an intrusion on their autonomy to lead a worthy self-directed life \citep{shiffrin2011thinker, baker1982process}. On this view, counterspeech is good simply because it is speech---it expresses the counter-speaker's views---and it is even better if it is embraced as an alternative to restricting the first speaker's speech.

Put this way, however, the autonomy (or ``liberty'') theory starts to run into serious difficulties. For one thing, it says little about the extent to which it is sufficient simply to speak, or whether the speaker's autonomy also requires that they reach an audience, and that the audience cooperate with them in their goals. If the former, this is a very weak right indeed; if the latter, it starts to intrude on others' autonomy. Effective counterspeech might mean that the original speaker \textit{fails} in their goal of persuading listeners---but the counterspeaker has just as strong a claim to reach an audience and persuade them, creating an insoluble conflict between speakers.

Another and more promising way of cashing out autonomy theories is that they are focused on listeners' autonomy in being able to freely decide for themselves, and that others are only allowed to affect their deliberations through persuasion rather than coercion \citep{scanlon1972}. On this view, censorship of speech is bad because it treats the listener as less than a fully rational being and deprives them of an opportunity to hear speech that they would find persuasive. But this also shows why counterspeech is so much better: it gives listeners a fair opportunity to hear \emph{multiple} points of view and autonomously make up their own minds among them.

\paragraph{Self-Government.}

Under the self-government theory, speech protections stem from participatory democracy \cite{weinstein2013}. For people to participate in democracy, they must have political autonomy---the ability to express and form their views on issues of public importance. The censorship of political speech violates this autonomy, both for speakers and listeners. It is worth noting that the self-government substantially overlaps with the search for truth theory, as suppression of political speech typically offends both \cite{volokh2013}.

Still, there are some important differences between the two, which have important implications for counterspeech. First, the self-government theory is driven by a particular suspicion of concentrated power, especially but not exclusively government power. Censorship is bad because it leads to \textit{domination} of the political process, and merely having the kind of power to shape public deliberation is still dangerous even when it is used benignly in a particular instance \citep{blasi1985pathological}. Thus, interventions like counterspeech are preferable because they limit the regulator's ability to intervene at all---a point that applies with equal force to governments and platforms.

Second, the self-government theory is more concerned with airing a well-structured range of views for public deliberation than with a wholly unfettered cacophony of speech. As Alexander Meiklejohn put it, in comparing a healthy speech environment to a New England town hall meeting, ``What is essential is not that everyone shall speak, but that everything worth saying shall be said'' \citep{meiklejohn1948free}. The relevance to counterspeech should be obvious; counterspeech is valuable to the extent that it adds to ``everything worth saying.'' Note also how listener-oriented this view is; the speech is valuable because listeners may find it valuable in their roles as citizens.

Certain regulations of speech illuminate the role of counterspeech under the self-government theory \citep{sunstein1992free}. The FCC's Fairness Doctrine is perhaps the canonical example, requiring broadcast television and radio networks to cover contrasting viewpoints on issues of public importance. The regulation reflects an aspirational vision under which counterspeech serves the project of self-government by providing access to a greater diversity of viewpoints. (Further discussion of the doctrine---particularly, in its practical limitations---is deferred to later.)

\subsection{Challenges to Counterspeech}

The assumption that counterspeech effectively combats misinformation has been heavily critiqued---especially as of late (e.g., \cite{napoli2018, tufekci2018, douek2021governing}). 
For our purposes, in considering potential interventions that promote counterspeech, it is useful to recount these critiques---not as reasons to abandon the counterspeech doctrine, but rather to clarify the challenges that any effective intervention must overcome. In this section, we consider why and when counterspeech struggles to correct misinformation, especially on social media. 

\paragraph{Scale and Speed.} 
The first challenge concerns the production of counterspeech to meet the scale of misinformation. While a high barrier of entry may have previously prevented the easy dissemination of misinformation, lower barriers of entry---while bringing greater diversity and democratization of speech \citep{volokh1994cheap}---allow misinformation to compete freely. In the open market, it appears that misinformation has a competitive advantage. 

Producing fake news is cheap \citep{condliffe2017fake}; unlike high-quality journalism, the creation of misinformation does not require the same resources needed in reporting. Technological advances including the emergence of generative AI have only further lowered the cost to producing misinformation \citep{citron2019deepfakes}. Moreover, fake news receives more attention, and travels faster. The folk wisdom that ``A Lie would travel from Maine to Georgia while Truth was getting on his boots'' still applies today (and perhaps more so). Research demonstrates that on social media, false news spreads faster than true news \citep{vosoughi2018}.

The scale of misinformation poses a basic challenge for counterspeech: to be effective, counterspeech must meet the scale and speed at which misinformation spreads.

\paragraph{Reaching the Audience.}
Even if enough counterspeech is produced, counterspeech still faces the challenge of reaching its audience. If a person is convinced by a piece of misinformation, but never sees the corrective counterspeech, then the counterspeech is ineffective. This issue is longstanding and commonplace.\footnote{For example, in their 2000 article, Robert D. Richards and Clay Calvert examine several examples of effective counterspeech, and conclude that ``counterspeech is most effective when its proponents are able to call journalistic attention to their message, place it on the media’s agenda, and thereby exponentially increase the audience to whom the message is disseminated'' \cite{richards2000counterspeech}.}

While the challenge of reaching the requisite audience has long diminished the efficacy of counterspeech, Phillip Napoli suggests that ``the ability of counterspeech to reach exactly those it needs to reach has been diminished as a result of the technological changes that have affected the media ecosystem.'' A primary reason is the emergence of ``filter bubbles'' on social media; if people are only shown content that agrees with their existing ideological beliefs, this implies that those who see misinformation are unlikely to see corrective counterspeech. 

\paragraph{Persuading the Audience.}

An issue remains, even when counterspeech is both produced and reaches the appropriate audience: When presented with corrective information, people do not always revise their beliefs. In extreme cases, presenting corrective information can cause a ``backfire effect,'' in which a person’s false belief is strengthened \citep{nyhan2010when}. Even if the backfire effect is relatively rare \citep{wood2019elusive}, there is growing consensus that corrections (e.g., fact checks) often have limited efficacy \citep{walter2019a, walter2019b}.

The inefficacy of counterspeech in correcting beliefs has been explained by the theory of motivated reasoning. The classical formulation developed in psychology posits that people have both \textit{accuracy} and \textit{directional} motives when processing information \citep{kunda1990}. People are both motivated by a desire to be accurate, but also by a desire to reach certain conclusions. The desire to reach particular conclusions is particularly present for political issues, in which case, arguments congruent with existing beliefs are evaluated as stronger \citep{taber2006}. In other words, people are less likely to incorporate correctional information when it runs against their ideological lean.

\section{Counterspeech Mechanisms}\label{sec:precedents}

Having established the role of counterspeech under different theories of free speech, as well as ways in which counterspeech falls short in achieving its purposes, we turn to the possibility of facilitating more effective counterspeech.

We conceptualize this task as a \textbf{counterspeech mechanism}: a way in which to select counterspeech to promote. The bridging-based approach we present is an example of a counterspeech mechanism, in which counterspeech is selected on the basis of diverse approval. In this section, we consider three existing counterspeech mechanisms---social media comments, fact-checking, and the fairness doctrine---and analyze how they select counterspeech and why these approaches fall short.

\subsection{Precedents}

\paragraph{Social media comments.}

Social media already offers ways to directly reply to other speech. The most basic example is the comment functionality prevalent on social media platforms. At face, this functionality represents the most \textit{laissez-faire} counterspeech, as it allows any type of speech to receive any type of counterspeech. In practice, due to the large quantity of comments (some by bots), platforms must select specific comments to promote---in essence, making a choice on how to select counterspeech. While approaches can vary, most platforms aim to maximize metrics such as popularity, relevance, and engagement.\footnote{https://transparency.meta.com/features/explaining-ranking/fb-feed-recommendations/}

These metrics do not clearly implement the purposes of counterspeech. For example, a post that contains false information may nonetheless receive supportive comments that receive widespread approval from users aligned with the ideological message of the post and those comments. Alternatively, a post may receive primarily negative (and perhaps corrective) comments from ideological opposite users, but these comments may be vitriolic in a way that fails to effectively persuade. Effective counterspeech does not appear to be competitive in a popularity contest.

\paragraph{Fact-checking.}
Fact checking is a counterspeech mechanism that more directly implements the purpose of counterspeech under a search-of-truth theory. Here, the speech that deserves counterspeech is misleading or false speech, and the counterspeech is written by a fact checker. For example, Meta's fact checking program surfaces potential misinformation using user feedback (a user can flag a post as false), which a third-party fact checker from a pre-chosen set of organizations then reviews. Meta can then attach a label to the content as potentially false and include corrective information from the fact check.\footnote{https://www.facebook.com/formedia/mjp/programs/third-party-fact-checking}

The discretion for what constitutes misinformation, however, rests with independent fact checkers. While this likely remains preferable over platforms themselves deciding, it may still clash with aspects of free speech theory skeptical of the fallibility or motivations of individuals and institutions in separating true from false.

Fact checking can also face challenges in its persuasive effect. As discussed earlier, correctional information does not always succeed at correcting false beliefs, especially when the correction runs counter to a person's ideological lean \citep{walter2019b}. The efficacy of fact-checking is further hampered by perceptions of independent fact checkers as biased \citep{flamini2019}.

Another issue is of scale and cost: while Meta may be able to invest hundreds of millions of dollars into fact checking programs, other platforms may not be able to. The cost of supporting such services also makes the approach vulnerable to cost-cutting measures \citep{huang2024, hsu2023}.
This raises questions about the sustainability of fact checking as an intervention. Even when fact-checking is well-funded, it is unclear if it can reach the scale necessary to combat misinformation \citep{ap2024india}.

\paragraph{The fairness doctrine.}

While fact checking can be viewed as a direct implementation of counterspeech in its search-for-truth purpose, the self-government purpose is perhaps most directly implemented by the fairness doctrine. The 1949 FCC regulation mandated broadcast television and radio to cover issues of public importance and to present contrasting views on these issues. The FCC wrote that 
\begin{quote}
\small
    If, as we believe to be the case, the public interest is best served in a democracy through the ability of the people to hear expositions of the various positions taken by responsible groups and individuals on particular topics and to choose between them, it is evidence that broadcast licensees have an affirmative duty generally to encourage and implement the broadcast of all sides of controversial public issues ... \citep{fcc1949}.
\end{quote}

The FCC's justification notes the public interest rationale in ensuring that a wide range of viewpoints on public issues are available. This justification is echoed in \textit{Red Lion v. FCC}, in which the Supreme Court upheld the doctrine as constitutional \citep{redlion}.

The fairness doctrine, as a counterspeech mechanism, makes choices about what speech deserves counterspeech (speech about important or controversial public issues) as well as what counterspeech should be promoted (speech containing contrasting views on the issue). Implementing this objective in regulation proved challenging, however. In practice, the FCC struggled to find a clear standard by which to apply the regulation. For example, it was unclear what issues met the bar of public importance, and what contrasting views deserved air time, ultimately resulting in few successful litigations \citep{krattenmaker1985fairness}. (At the same time, the doctrine was susceptible to use by political actors to chill opposing speech, since litigation could be expensive \citep{cronauer1994fairness}. Furthermore, there is evidence that the doctrine led to a chilling effect on coverage of controversial public issues, due to broadcast stations avoiding issues that may trigger the doctrine \citep{hazlett1997was}.)

The implementation of the fairness doctrine revealed a lack of workable standards and a high level of discretion. So even as recent concerns about platform fairness have brought the doctrine back into conversation \citep{napoli2021back}, by the measures considered here, it is unlikely that a fairness-doctrine-like mechanism would be an adequate counterspeech mechanism on social media.

\begin{table}
\centering
\begin{tabular}{l l}
\toprule
\textbf{Mechanism} & \textbf{Counterspeech Selection Rule} \\
\midrule
Social media comments & Most popular, engaging, or relevant\\
Fact-checking & Corrects misinformation \\
Fairness doctrine & Opposing viewpoint \\
\midrule
Bridging & Approval from diverse viewpoints \\
\bottomrule
\end{tabular}
\caption{Counterspeech mechanisms}
\end{table}

\section{A Bridging-Based Approach}\label{sec:bridging}
Here, we argue that a bridging-based approach to selecting counterspeech overcomes many of the present challenges, as well as shortcomings of past counterspeech mechanisms. As a counterspeech mechanism, the bridging-based approach promotes counterspeech that is \textit{approved by people of diverse viewpoints}. This objective has been proposed as a general alternative to engagement optimization, which have been shown to be misaligned with user and societal goals \citep{kleinberg2024challenge, milli2023engagement}. ``Bridging-based ranking,'' on the other hand, aims to bridge divides \citep{ovadya2023bridging}. Here, we argue that the objective is particularly suitable for promoting counterspeech to combat misinformation.

In this section, we begin by discussing why approval by diverse viewpoints is the right objective. First, we show that diverse approval selects counterspeech that is corrective in its effect---that is, it provides information from which people might actually update their beliefs. In comparison to fact checking, for example, diverse approval ensures that the counterspeech not only corrects the original claim, but is also well-received by the individuals who were most likely to believe the misinformation to begin with. This shows bridging's theoretical appeal. Many of bridging's strengths, however, lie in its implementation. The algorithmic approach, which we discuss in some detail, operates through crowd-sourced voting. This allows bridging to operate at scale, and without overt platform or government discretion. The algorithm is concrete, and content- and viewpoint-neutral---allowing the objective to maintain its integrity in practice (\textit{contra} the fairness doctrine).

An initial version of the bridging-based approach has been implemented on Twitter/X in recent years, providing a concrete setting to analyze. We describe this setting in detail, and discuss the initial evidence on its success. The setting demonstrates the strength of the approach (e.g., providing fact checks that are accurate and persuasive), while also highlighting potential challenges (e.g., building consensus before the misinformation spreads too widely). We conclude the section by discussing critiques and challenges of the approach, while highlighting important directions for future research.

\subsection{Why is bridging the right mechanism?}

Obtaining approval from diverse viewpoints is not obviously the right approach to correcting misinformation. One could alternatively try to simply identify misinformation and then correct it---this being the approach traditional fact checking takes. Here, we make the argument that bridging is a better method. Most centrally, we argue that bridging is simply more effective at identifying effective counterspeech. However, bridging also satisfies numerous other desiderata where other mechanisms fall short. The algorithmic implementation of bridging shifts discretion from platforms to users, enables production at scale, has a concrete content- and viewpoint-neutral objective, and is in alignment with listener (i.e., user) goals, as well as platform goals.

We now argue that bridging is necessary to ensure that counterspeech is corrective in its effect. Consider the following simple conceptual model. Consider a piece of misinformation, and two groups of people, $A$ and $B$. Group $A$ is inclined to be skeptical of the misinformation, while group $B$ tends to support it. (In practice, these groups often map onto the ideological spectrum.) For example, regarding speech that argues that climate change doesn't exist, group $A$ is likely to contain more left-leaning individuals and group $B$ is likely to contain more right-leaning individuals.

Now consider counterspeech that does not obtain approval across these two groups. We argue that such counterspeech is unlikely to be effective in correcting misinformation. Counterspeech that is only supported by group $B$ is unlikely to be corrective in its content, but will likely instead amplify the original speech. Therefore, it is unlikely to be effective counterspeech, since it simply doesn't aim to correct the original claim. On the other hand, counterspeech that is only supported by group $A$ is likely to be corrective (i.e., serve what we more traditionally think of as counterspeech or fact checking). However, if the speech is not approved of by individuals from group $B$, then the counterspeech remains ineffective: it does not correct the beliefs of the group most likely to believe the misinformation. Counterspeech that is supported by both groups $A$ and $B$ is thus more likely to be corrective, while also being received and incorporated by those who might be misled by the original speech.

In this way, the bridging-based objective attempts to directly overcome fact checking's struggle to persuade. The bridging-based objective does this by asking the people directly: did you find this counterspeech to be useful? Whereas fact checking can provide corrective information, it does not directly ensure that this corrective information will change minds. The ability of bridging to produce effective fact checks is---by itself---a compelling reason to support the bridging-based approach.

This model also explains the empirical observation that the ``demand-driven'' approach to selecting counterspeech is unsuccessful: popular counterspeech often fails to be bridging, and therefore fails to be corrective in its effect. Critics of the \textit{laissez-faire}, marketplace theory point to this fact as reason to support more direct content moderation or fact checking, while supporters of the theory hang on to the belief that platform discretion is the worse of two evils. The model shows how bridging reconciles this tension: bridging is an objective that depends on the preferences of people, not the platform, while also being better at correcting misinformation than direct platform discretion.

To be truly convinced of bridging, one must consider its implementation. The history of the fairness doctrine, for example, demonstrates that even when an objective is desirable in theory, implementation can significantly diminish its desirability. Before describing bridging's implementation in detail, we highlight the implementation's key property. The algorithmic implementation of bridging aggregates user votes into a score; if the score exceeds a threshold, the counterspeech is displayed. The crowd-sourced nature means that the approach shifts discretion from platforms to users, which consequently also enables the mechanism to operate at scale, overcoming a primary hurdle presented by the modern media environment. Moreover, the algorithm is transparent and content- and viewpoint-neutral, overcoming the challenges the fairness doctrine faced in practice.

\subsection{An algorithmic implementation}

How can we identifying counterspeech that is bridging---that is, counterspeech that obtains approval from diverse viewpoints. Here, we describe an algorithmic implementation of bridging that builds upon the \textit{collaborative filtering} approach that forms the foundation of all modern social media.

The fundamental challenge of bridging is identifying users of diverse viewpoints. Once we have this, one can examine the votes of different users to identify counterspeech that is bridging. The key idea is to simultaneously use users' voting behavior to understand their viewpoints. Intuitively, users with different voting behavior are more likely to have different viewpoints. This is the same intuition that underlies collaborative filtering. Consider, for example, the task of recommending movies. By looking at all users' watch histories, it is possible to identify users with similar preferences. This is useful for a movie-streaming platform, since it can then recommend movies that ``users like you also watched.'' Indeed, in 2007, Netflix offered a one million dollar prize to the team that could best solve this collaborative filtering task \citep{netflixprize}.

Collaborative filtering methods can be adopted to implement bridging. Here, we describe the algorithmic implementation of bridging used on Twitter/X's Community Notes program. (We further discuss Community Notes in the next section.) The Community Notes algorithm uses a simple approach motivated by matrix factorization, a widely-used collaborative filtering method (see, e.g, \cite{koren2009matrix}). The approach simultaneously learns the ``viewpoint'' of users as well as whether or not a piece of counterspeech is bridging. The algorithm is described by \cite{wojcik2022birdwatch}. We describe a slightly simplified version below.

Consider a user $u$ and a note $n$ (a ``note'' is a piece of counterspeech being evaluated). Then each user has two attributes $i_u$ and $f_u$; each note likewise has two attributes $i_n$ and $f_n$. We learn these attributes using the voting data between users and notes. In particular, for a note $n$, we predict that user $u$ will vote
\begin{equation}\label{eq:cn-prediction}
    \hat{r}_{un} = i_u + i_n + f_u\cdot f_n.
\end{equation}
The goal is to choose $i_u, f_u, i_n, f_n$ for each user and note to minimize
\begin{equation}
    \sum_{(u,n)} (\hat{r}_{un} - r_{un})^2,
\end{equation}
where $r_{un}$ is how user $u$ actually voted on note $n$ ($1$ representing approval, and $-1$ disapproval). This can be done with standard machine learning techniques. We then score a note based on the value of $i_n$ (higher is better).

The intuition is as follows. $i_u$ represents how likely a user is to approve of a note in general. $i_n$ represents how likely a note is to be approved of by a user in general. Therefore, both higher $i_u$ and $i_n$ result in a higher prediction in \eqref{eq:cn-prediction}. Meanwhile, $f_u$ and $f_n$ reflect the ``viewpoints'' of the user and the note. If $f_u$ and $f_n$ are aligned (share the same sign), then this also results in a higher prediction. For example, if $f_u=f_n=-1,$ then $f_u\cdot f_n=1$; if $f_u=1, f_n=-1,$ then $f_u\cdot f_n = -1.$ Intuitively, if the user shares the same viewpoint as the note, the user is more likely to approve. This means that when $|f_n|$ is large, voting behavior varies a lot by viewpoint. Therefore, if a note is not bridging, meaning that users of different viewpoints vote differently, this will result in a significant $f_n$ term but a small $i_n$ term. Meanwhile, if a note is bridging, meaning that users of different viewpoints vote positively, then this voting behavior is best explained by a high $i_n$ and an insignificant $f_n$.

The Community Notes algorithm illustrates how it is possible to implement bridging using only the voting behavior of users. This enables a crowd-sourcing approach that can enable production of counterspeech at scale, while giving discretion to users and not the platform. Note also that the algorithm does not explicitly evaluate based on any predetermined viewpoints (e.g., ``liberal'' or ``conservative''), but rather learns viewpoints to be the characteristics that most explain variation in how users vote. This approach can also be implemented in a transparent way; the Community Notes algorithm's decisions can be reproduced, for example, using publicly available code and data.\footnote{\url{https://github.com/twitter/communitynotes}}

\subsection{Case study: Community Notes}

The bridging-based approach has been implemented through X's Community Notes program \citep{wojcik2022birdwatch}. This provides us a setting where we can evaluate the potential efficacy of the approach in practice. The basic functionality of Community Notes is simple, as described by X:
\begin{quote}
    Community Notes aim to create a better informed world by empowering people on X to collaboratively add context to potentially misleading posts. Contributors can leave notes on any post and if enough contributors from different points of view rate that note as helpful, the note will be publicly shown on a post.\footnote{\url{https://help.twitter.com/en/using-x/community-notes}}
\end{quote}
Here, a \textit{note} is a piece of counterspeech that responds to a \textit{post}, where the intention is for the counterspeech to provide context to misleading posts. The goal of the program, thus, is to help combat misinformation. The \textit{mechanism} by which Community Notes implements this goal is through the bridging-based approach described in the previous section; as described by X, a note is made public if ``enough contributors from different points of view rate that note as helpful.''

For concreteness, we provide three examples of notes that have been made public through the program. These examples illustrate how notes can correct different kinds of misinformation. In the first example (Figure 1), U.S. Representative Lauren Boebert implicitly claims credit for legislation that advanced local priorities. The note adds context that Boebert voted against the referenced legislation. The Community Note provides a link to the government record of the vote. The note also contains a qualification, that ``Congresswoman Boebert may have worked with local stakeholders,'' making clear the specific part of Boebert's post that the note refutes. The second example (Figure 2) is a post by the official White House account during the Biden administration, which claims that ``750,000 manufacturing jobs have been created under Biden's leadership.'' The Community Note adds context that the number is true compared to 2021, but that the number is smaller (150k) when compared to 2020. Therefore, a large number of these added jobs are a product of the pandemic recovery. The note provides a link to government data reflecting these numbers. The third example (Figure 3) is a post by a user with more than 2 million followers on the platform. The post refers to an article that notes a recent increase in early-onset cancer, adding: ``I'm stumped here. Can't think of anything different these last few years. Can you?'' The sarcastic rhetorical question draws a connection between the adoption of COVID-19 vaccines and the increasing cancer rates. The Community Note provides context that that the study mentioned by the article covers years up to 2019, \textit{before} the introduction of the COVID-19 vaccines.

These examples illustrate the potential capabilities of Community Notes and bridging-based counterspeech interventions. First, all three examples deal with posts that are misleading in fairly subtle ways. None of the posts \textit{directly} make claims that are clearly false. The post by the Biden White House is not false, but the additional context makes it possible to evaluate the claimed that Biden's leadership significantly increased the number of manufacturing jobs. Meanwhile, the post by Joey Barton never even mentions vaccines---yet, the Community Note addresses the implied connection. Second, the examples illustrate that Community Notes are used to add context to posts coming from across the ideological spectrum, as demonstrated by Community Notes being added to both representative Boebert (a Republican), and Biden's White House (Democratic). Third, the last example demonstrates how Community Notes can be useful even for highly-controversial topics, such as vaccines. In particular, it appears possible for notes to achieve diverse approval in these controversial settings, despite skepticism (see, e.g., \cite{czopek2023}).

An emerging body of work has examined the efficacy of Community Notes in practice.
Internal research by X, based on surveys, suggests that users who see a note are 20-40\% less likely to believe the substance of an annotated post than those who did not see the note \citep{x2022birdwatch}. External research also suggests that Community Notes reduces the spread of misinformation, though the intervention is less effective of curtailing spread early on \citep{chuai2024}. Other work has documented the extent to which Community Notes produces accurate fact checks on vaccine misinformation, finding that more than 97 percent of vaccine-related Community Notes were accurate, and were viewed more than 200 million times \citep{allen2024vaccine}.

\subsection{Critiques and Concerns}

We now consider three critiques of the bridging-based approach, and discussing the needed future research in these directions.

\paragraph{Overly restrictive selection criteria.}
One weakness of the bridging-based approach is that it is restrictive in the counterspeech that it ultimately promotes. Because the approach requires that counterspeech receive approval across a diverse audience, this necessarily limits some counterspeech. Alex Mahadevan, the director of the Poynter Institute's MediaWise, said, ``So this algorithm that was supposed to solve the problem of biased fact-checkers basically means there is no fact-checking'' \citep{czopek2023}. Mahadevan raises the concern that the high bar for counterspeech means that little fact-checking can occur. However, an examination of the type of counterspeech that is shown on Community Notes reveals that a significant amount of counterspeech reaches this bar, even on controversial topics such as vaccines and COVID-19. Moreover, in the argument we have laid out, bridging is to some degree necessary to ensure that counterspeech is effective in convincing users who are persuaded by the misinformation. More research is needed to understand the when, and to what degree, the bridging-based approach is limited in accomplishing fact-checking.

\paragraph{Speed and scale.}

Initial evidence suggests that ``the half-life of reposts over 36 hours is 5.75 hours with only 13.5\% of all helpful notes displayed before this time point'' \citep{chuai2024}. This means that the efficacy of Community Notes was relatively small, as measured by limiting the number of times a post was shared; effective Community Notes were not created and voted on fast enough to be shown in the early stage of misinformation. The efficacy of Community Notes, however improved over time in the months after it was first introduced \citep{chuai2024}. This suggests that as the program is more widely adopted, issues of speed may be overcome. The challenge of promoting counterspeech quickly introduces several directions for future work in algorithm design, exploring whether modifying the algorithm can reduce the number of votes to reach consensus, or if there is a way to nudge users into voting on certain pieces of counterspeech.

\paragraph{Gameability.}

Another concern, as is common in crowd-sourced approaches, is the gameability of the mechanism. However, the bridging-based objective, as implemented through Community Notes, appears fairly robust to strategic behavior. It is possible that this is due to the relatively limited pilot nature of Community Notes, where voting is still limited to a set of (relatively) early adopters. This group may, as a whole, be more aligned with broader community interests. However, the degree to which strategic individuals or groups can co-opt the program to display biased messages or take down notes attached to messages they find favorable is not yet clear. Further research would provide greater insight into this concern.

 \bibliographystyle{ACM-Reference-Format}
 \balance
 \bibliography{bib}

\newpage
\appendix

\onecolumn

\begin{figure}
\begin{center}
    \includegraphics[width=0.5\linewidth]{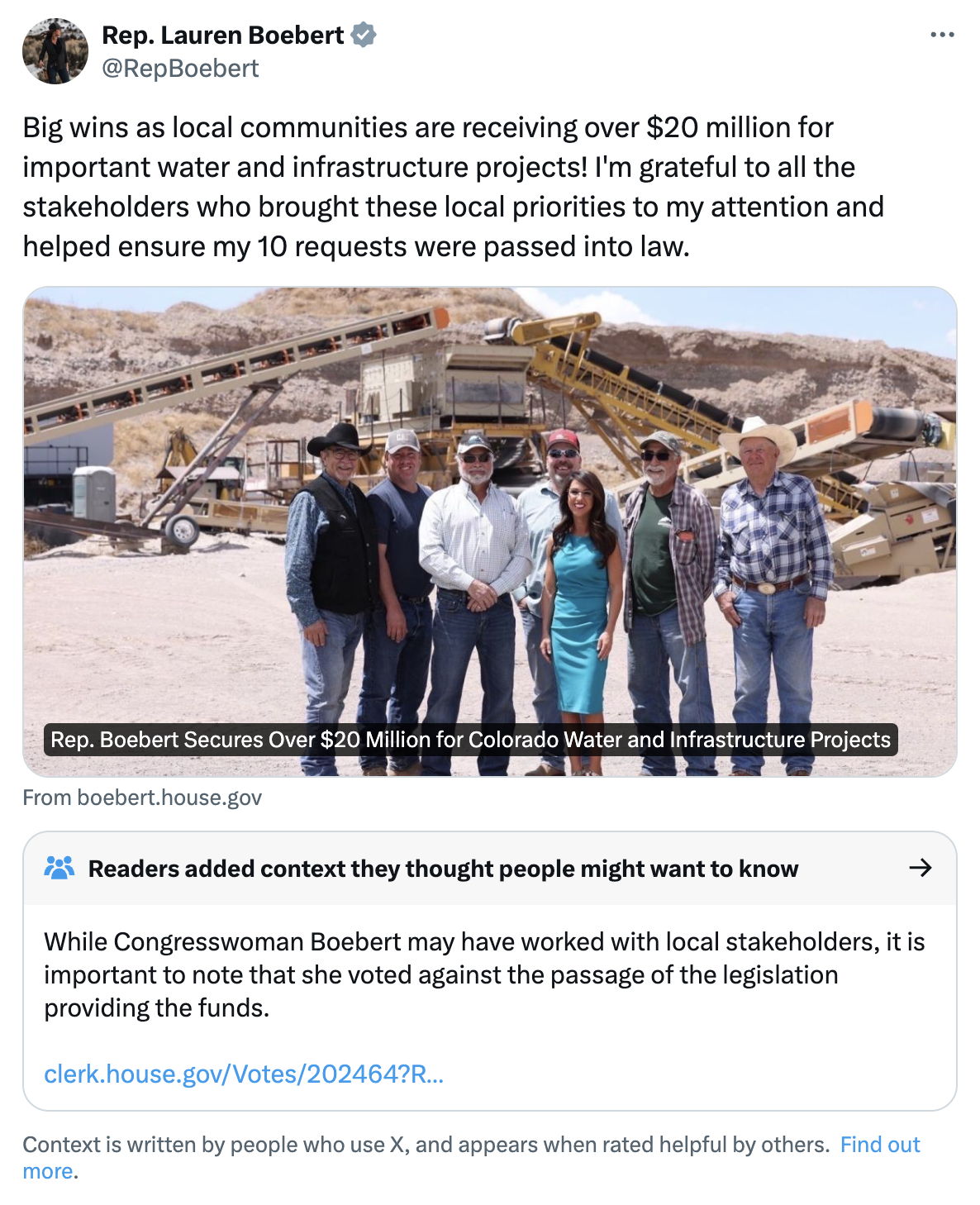}
\end{center}
\caption{An example of a Community Note added to a post by U.S. Representative Lauren Boebert.}
\label{fig:boebert}
\end{figure}

\begin{figure}
\begin{center}
    \includegraphics[width=0.5\linewidth]{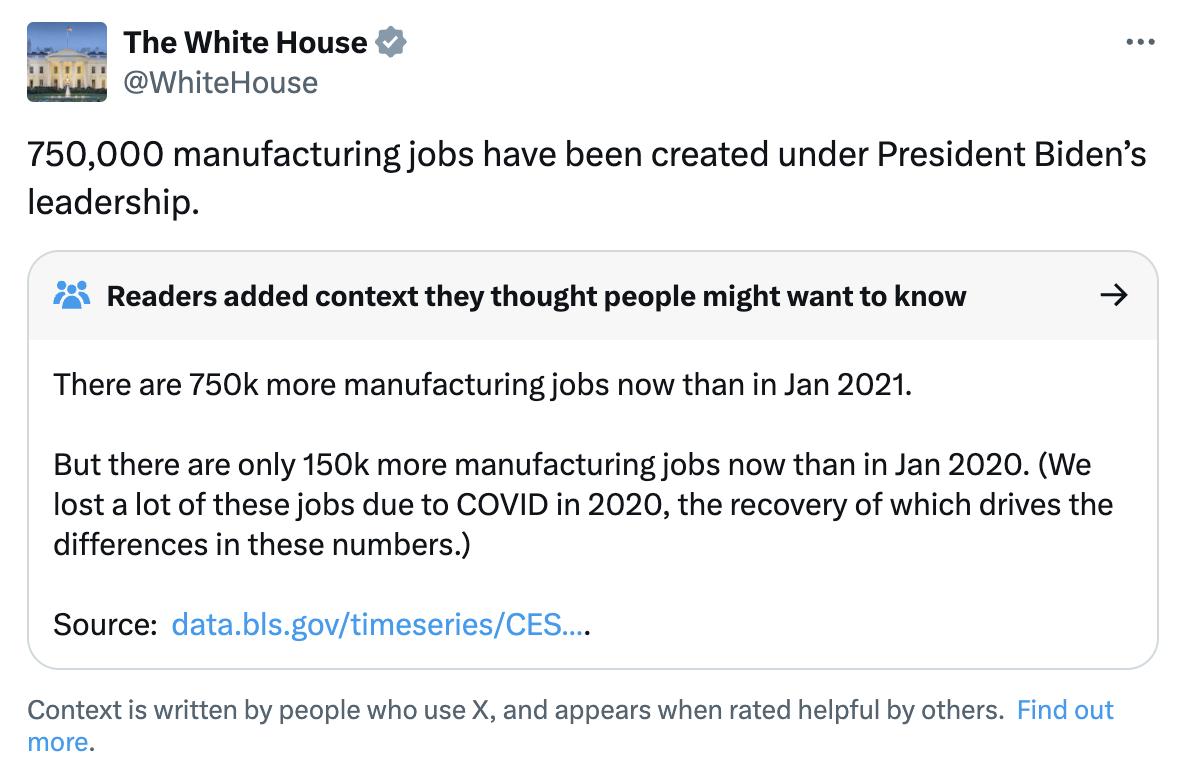}
\end{center}
\caption{An example of a Community Note added to a post by the Biden White House.}
\label{fig:white-house}
\end{figure}

\begin{figure}
\begin{center}
    \includegraphics[width=0.5\linewidth]{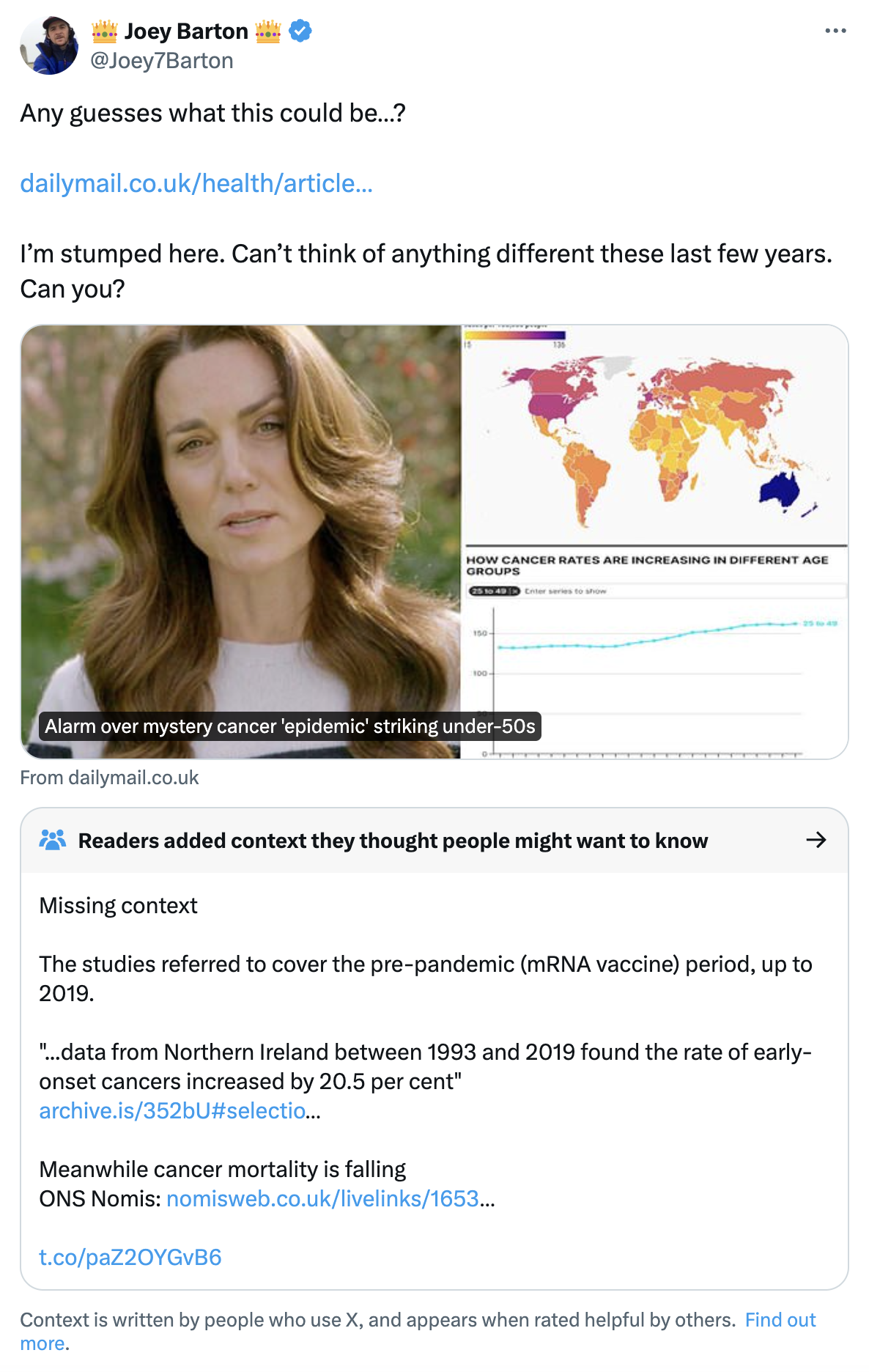}
\end{center}
\caption{An example of a Community Note added to a post by a widely-followed user promoting vaccine misinformation.}
\label{fig:vaccine}
\end{figure}

\end{document}